# Photoconductivity of Single-crystalline Selenium Nanotubes

Peng Liu[1], Yurong Ma[2], Weiwei Cai[1], Zhenzhong Wang[1], Limin Qi[2] and Dongmin Chen[1]

[1] Beijing National Laboratory for Condensed Matter Physics, Institute of Physics, Chinese Academy of Sciences, Beijing 100080, China

[2] State Key Laboratory for Structural Chemistry of Unstable and Stable Species, College of Chemistry, Peking University, Beijing 100871, China

E-mail: liup@aphy.iphy.ac.cn

## Abstract

Photoconductivity of single-crystalline selenium nanotubes (SCSNT) under a range of illumination intensities of a 633nm laser is carried out with a novel two terminal device arrangement at room temperature. It's found that SCSNT forms Schottky barriers with the W and Au contacts, and the barrier height is a function of the light intensities. In low illumination regime below $1.46\times10^{-4}\,\mu W\mu m^{-2}$, the Au-Se-W hybrid structure exhibits sharp switch on/off behavior, and the turn-on voltages decrease with increasing illuminating intensities. In the high illumination regime above $7\times10^{-4}\,\mu W\mu m^{-2}$, the device exhibits ohmic conductance with a photoconductivity as high as $0.59\Omega cm^{-1}$, significantly higher that reported values for carbon and GaN nanotubes. This finding suggests that SCSNT is potentially a good photo-sensor material as well we a very effective solar cell material.

## Introduction

As an important elemental semiconductor, selenium shows a variety of interesting properties, such as high photoconductivity, nonlinear optical response, and has commercial applications in photovoltaic cells, rectifiers, photographic exposure meters, and xerography. Like other photonic semiconductor [1-4], when the dimension and sizes are reduced, selenium nanostructure is expected to show some quantum-size effects that might offer new or improved photonic applications. In recent years, some groups have succeeded in developing new synthetic methods to fabricate 1D selenium nanostructures [5, 6]. So far, however, there is still very limited report on the physical properties of the nanostructure of Se, especially as an electrical or photonic device. Here we report on a first detail study of the photo conductance of trigonal selenium nanotubes fabricated by a unique, facile and large-scale synthesis method. Extending the scanning tunneling microscopy (STM) stepper techniques [7], we have successfully developed a novel approach to making clean electrical contacts to individual Se nanotube to form two-terminal devices for our transport measurement. This versatile technique avoids contact problems often encountered in lithographically patterned devices due to contamination or damage from energetic electron or ion beams. The device made using present technique form reliable Schottky barrier at the Semiconductor-metal contacts and a back-to-back Schottky diode device. Together with the photo excitation of the carrier under different light intensities, this type of device exhibits I-V characteristics suitable for photo-sensor and photo-cell applications.

**Experimental details**

The single-crystalline t-Se nanotubes were synthesized by the dismutation of $Na_2SeSO_3$ under acidic condition in micellar solutions of the surfactant poly (oxyethylene) dodecyl ether $C_{12}E_{23}$, which is a non-ionic surfactant with a low critical micelle concentration (cmc) in aqueous solution (typically $4 \times 10^{-5}$-$2 \times 10^{-4}$M) [6]. Figure 1 shows scanning electron microscope (SEM) image of the t-Se nanotubes. The nanotubes' diameters are typically ranging from 80 to 300nm and lengths ranging from several μ m to more than 100μ m. The tubes have a well-facetted prism morphology with a relatively uniform wall thickness (30-50nm) and pseudo-hexagonal or pseudo-triangular cross sections, indicating the formation of single-crystalline t-Se nanotubes with {110} planes on the sides [6].

To perform electrical transport measurement of single nanotube, we fetch individual tube out from the solution by means of dielectrophorisis (DEP). We prepared a sharp tungsten tip by electrochemical etching in KOH solution of 5mo1L$^{-1}$[8], followed by chemical etching with dilute HF solution to remove oxidized layers [9]. As illustrated in Figure 2(a), the tungsten tip is positioned above a metal electrode (a few hundreds nm thick Au film deposited on silicon dioxide substrate) via a high precision mechanical stage. The apex of tungsten tip was illuminated by a LED and monitored by a CCD camera in real-time. The tip-facing-tip image in the upper-right of Figure 2 (a) was the CCD image (with 100 times magnification). From this image, we could estimate the distance between the tip apex and electrode was about 30μm. After fixing the tip at this position, we dropped nanotubes solution to the apex part, and turned on the function generator to generate an AC electrical field. In this process, nanotubes and nanoparticles with longer dimension having larger dipole moment, will be attracted and aligned to the tip sooner than small particles, such as impurities [10]. Figure 1 (b) shows a typical image of a SCSNT attached to the tip using this DEP process.

Our two terminal transport measurement system is shown in Fig. 2b, where the nanotube forms one contact with the tip, and the other with the Au electrode. A computer-controlled piezo-driven stepper is used to move the tip and nanotube towards the Au electrode at a 20nm step and about 20 steps per second [7]. A feedback circuit similar to a scanning tunneling microscope servo system is used to promptly hold the tip position when a current is detected as a result of the contact between the nanotube and the Au electrode. This method leads to reproducible and controllable contacts and avoid the oxidation and contamination problems often encountered in the conventional patterned electrode using the lithographical technique.

## Results and discussion

For photoconductivity measurements, we illuminate the SCSNT with a He-Ne laser ($\lambda$ = 633nm), and the illumination intensity was precisely controlled by two polarizers. One polarizer is parallel to laser's polarized axis while the other is rotated with respect to the first polarizer to yield the desired light throughput. The laser spot was about 650μm in diameter and its maximum power was 2mW in the central part. Assuming a Gaussian distribution of the laser beam and neglecting the scattering from the SCSNT, we estimate that the maximum area power near the center of the SCSNT was about $60 \times 10^{-4}$ μWμm$^{-2}$.

We first measured the dark current using an aluminum foil to shield the ambient light. As shown in the black curve in Fig 3a, under dark condition the I-V exhibits a gap of 2.24eV, with a characteristic of two back-to-back Schottky diodes in the W-Se-Au structure. The reverse bias break down occur at +0.99 and -1.25 volt, respectively. Note that the gain of the current preamplifier was set to $10^8$, so that the current saturate at $\pm$100nA.

The photocondutance for various illumination intensities is shown in Fig. 3a-3c. From $0.018 \times 10^{-4}$ to $1.46 \times 10^{-4}$ μWμm$^{-2}$ illumination(Fig. 3a), the *I-V* spectra continues to show a reverse bias break though character at both positive and negative biases, but the gap reduces as the photo power increases, suggesting a reduction of the Schottky barrier heights at the nanotube-metal contacts. Note that there is slight asymmetry on the barrier height at these contacts, which can be attributed to the difference in the work function between W-tip and Au-electrode. When the illumination power rose to between $1.5 \times 10^{-4}$ μWμm$^{-2}$ and $5.7 \times 10^{-4}$ μWμm$^{-2}$, the sharp break down like I-V curves are now dominated by reverse leakage current for either biases as shown in Fig. 3b. Finally, when the illumination power exceeds $7 \times 10^{-4}$ μWμm$^{-2}$, the *I(V)* curves show ohmic behavior(Fig.3c). Beyond $25 \times 10^{-4}$ μWμm$^{-2}$ photo power, the slope of the I-V curve does not change any further, indicating that the carrier saturation is reached.

Trigonal selenium is generally accepted as a p-type extrinsic semiconductor, and conduction occurs due to valence band hole transport [11]. Despite the fact that trigonal selenium has a band gap of about 1.6eV, the room-temperature dark conductivity is usually in the range of $10^{-6}$ - $10^{-5}\Omega cm^{-1}$. Thermoelectric power measurements indicate that at room-temperature the majority carriers' concentration is about $10^{13} cm^{-3}$ [12]. The bulk material electron work functions of W, Se, and Au are 4.55eV, 5.9eV and 5.1eV, respectively. Under equilibrium condition at zero bias, metal-semiconductor(MS) contact and charge transfer cause the band bending near the contacts and Schottky barrier formation as illustrated in Fig. 4a, resulting in a back-to-back diode device of Fig. 4b. When a voltage was applied across the two MS contacts, one diode is forward biased while the other is reversed biased as shown in Figure 4(c) and 4(d). For example, when a positive voltage is applied to the W tip, W-Se Schottky junction was reverse biased while Au-Se junction was forward biased. In low bias regime (0.99V>V>-1.25V), the circuit was in an OFF-state. The applied voltage on W tip $V_A = V_{W-Se} + V_{Au-Se}$. When $V_A$ equates $V_{Flat-Band}$ [13], most of $V_A$ (positive biased for example) was applied on W-Se MS junction. When $V_A$ was raised to $V_{Breakdown}$ =1.25V in dark condition, reverse avalanche breakdown occurred in W-Se junction [13], and there is a rapid rise of the current.  The same happens to the Au-Se junction for negative polarity but the turn-on voltage is slightly different due to the variation of the work-function difference.

Under illumination, photo-generated carrier will raise the Fermi level in the Se nanotube and hence lower the Schottky barriers at the MS contacts. This results in a lower switch-on bias voltage as shown in Fig, 3a. As the barrier height lowers further with increasing illumination intensity, the reverse breakdown voltage was not enough to induce the avalanche process, so the current rises slowly instead (Fig. 3b). Since most of the bias voltage falls at the reverse bias junction, the switch on points are the measure of the respective Schottky barrier heights. Figure 5a plots the change of the Schottky barrier heights as a function of illumination intensity. It shows that the Schottky barriers decreased continuously with increasing illumination intensity but the difference between the Schoottky barrier heights is almost constant of ~ 0.26eV (within the measurement error bar) before reaching carrier saturation. It should be noted that the illumination intensity values is somewhat over estimated because we have not taken into account the effects of the incident light scattering from the Se nanotube.

When the illumination intensity equal and exceeds $7\times10^{-4} \mu W\mu m^{-2}$, both Schottky barriers nearly disappeared and the conductance exhibits ohmic behavior as shown in Fig. 3c. Finally, when the illumination power reaches $25 \times 10^{-4}$ µWµm$^{-2}$ and above the conductance is saturated. Fig. 5b (Add this figure) shows the conductance as a function of the illumination intensity where σ = L·I/V·S is extracted from the linear fits of the data in Fig. 3a with L=10µm, φ=320nm, and wall thickness=50nm. Note that R = R$_{Au-Se}$ + R$_{W-Se}$ + R$_{se}$, so the saturation might indicate that the contact resistances are the limiting factor and the photo conductance of Se nanotube could be much higher than 0.59 (Ωcm)$^{-1}$.

The dependence of photoconductivity on the illumination intensity is mainly determined by the recombination and trapping of the electron-hole pairs within solid materials, and the rate of such a recombination and trapping for selenium has been shown to strongly depend on the temperature [5]. Both light adsorption and resistive heating can raise the sample temperature and affect the conductance. In our study, however, the illumination power was so low that its heating effect should not have a significant influence on the photoconductivity. Previous work on phtoconductivity of selenium focused on amorphous selenium, liquid selenium and hexagonal metallic selenium film or bulk materials [14]. A recent experiment on t-Se nanowire which yielded σ = 12.4 (Ωcm)$^{-1}$ with ~ µWµm$^{-2}$ illumination intensity [5]. Our results indicate that with 4 orders of lower illumination intensities, ~10$^{-4}$ µWµm$^{-2}$, our SCSNTs exhibit much higher photoconductivity of 0.59 (Ωcm)$^{-1}$. This also compares favorably with other nanomaterials, such as single-walled carbon nanotubes (kWcm$^{-2}$, 0.38 (Ωcm)$^{-1}$) [15], and GaN nanowires (15Wcm$^{-2}$, 0.026 (Ωcm)$^{-1}$) [16]. This unique property might lend t-Se some superior application such as photo sensor and photo cell materials.

## Conclusions

In conclusion, we have measured photoconductivity of SCSNT by a two-terminal measurement system under different illumination intensities at room temperature. In low illumination regime below $1.46 \times 10^{-4}$ µWµm$^{-2}$, the Au-Se-W hybrid structure exhibits sharp switch-on and switch-off behavior. And with the increase of illuminating intensities, the turn-on voltage values decrease. This finishing suggests that SCSNT is potentially a good photo-sensor material. On the other hand, in high illumination regime above $7 \times 10^{-4}$ µWµm$^{-2}$, due to high photo-generated carrier density, SCSNT exhibits exceptional high photoconductivity. This indicates that the SCSNT could also be a very effective solar cell material. Further study on the spectra dependence of the photo-conductivity of SCSNT is clearly of great interest.

## Acknowledgments

The authors would like to thank Zhi Xu and Xuedong Bai for their kind help with TEM, Fei Pang for his kind help with software programming, and Jian Wang for his fruitful and helpful discussion.

Financial suppot of this work by a grant (No. 50518002) from NSFC-HKRGC is greatfully acknowledged.

## References

[01] S. Frank, P. Poncharal, Z. L. Wang, W. A. de Heer, *Science* 280, 1744 (1998)

[02] J. Hu, T. W. Odom, C. Lieber, *Acc. Chem. Res.* 32, 435 (1999)

[03] Z. L. Wang, *Adv. Mater.* 12, 1295 (2000)

[04] Y. Xia , P. Yang, Y. Sun, Y. Wu, B. Mayers, B. Gates, Y. Yin, F. Kim, Y. Yan, *Adv. Mater.* 15, 353 (2003)

[05] B. Gates, B. Mayers, B. Cattle and Y. Xia, *Adv. Funct. Mater.* 12, 219 (2002)

[06] Y. Ma, L. Qi, J. Ma, and H. Cheng, *Adv. Mater.* 16, 1023 (2004)

[07] H. Okamoto, and D. M. Chen, *Rev. Sci. Inst.*, 72, 4398 (2001)

[08] V. V. Dremov, V. A. Makarenko, S. Y. Shapoval, O. V. Trofimov, V. G. Beshenkov, and I. I. Khodos, *Nanobiology* 3, 83 (1994)

[09] J. E. Fasth, B. Loberg, and H. Norden, *J. Sci. Instrum.* 44, 1044 (1967)

[10] H. W. Lee, S. H. Kim and Y. K. Kwak, *Rev. Sci. Inst.* 76, 046108 (2005)
In nonuniform electric field, the DEP force can be expressed by $F_{DEP} = 2\pi a^3 \varepsilon_m \text{Re}[\varepsilon_p^* - \varepsilon_m^* / \varepsilon_p^* + 2\varepsilon_m^*] \nabla |E|^2$, where $a$ is the longest dimension of the particle, $\varepsilon_m$ is the dielectric constant of the medium, $\varepsilon_p$ is the dielectric constant of the particle, and E is the electric field. When we set the AC voltage $V_{PP}$ = 15V and frequency $f$ = 1M Hz, the maximum electric field was about $2.5 \times 10^5 \text{Vm}^{-1}$.

[11] J. Mort, *Phys. Rev. Lett.* 18, 540 (1967)

[12] G. B. Abdullayev, N. Z. Dzhalilov, and G. M. Aliyev, *Phys. Lett.* 23, 217 (1966)

[13] M. Grundmann, *The Physics of Semiconductors*, Springer, **2006**, pp. 492

[14] T. S. Moss, *Photoconductivity in the Elements*, Butterworths, London **1952**, pp. 185-203

[15] M. Freitag, Y. Martin, J. A. Misewich, R. Martel and Ph. Avouris, *Nano Lett.*, 3, 1067 (2003)

[16] Raffaella Calarco, Michel Marso, Thomas Richter, Ali I. Aykanat, Ralph Meijers, Andre v.d. Hart, Toma Stoica, and Hans Luth, *Nano Lett.*, 5, 981 (2005)

**Figure Captions:**

**Figure 1**: (a) SEM image of single-crystalline Se nanotubes as synthesized; (b) TEM image of a SCSNT attached to a tungsten tip.

**Figure 2**: Schematics of (a) dielectrophorisis apparatus used to attach single SCSNT to a metal tip. and (b) a single SCSNT two-terminal device arrangement for photoconductivity measurement.

**Figure 3**: *I(V)* characteristics of SCSNT under different illumination intensities. The inset shows a equivalent device model of two back-to-back Schottky barrier diodes.

**Figure 4**: Energy band diagram of Au-Se-W device (a) in equilibrium state with zero bias and under dark condition; (b) under low bias condition showing bend bending; (c) under high bias condition showing reverse junction break-down; and (d) under light illumination which reduces the barrier heights.

**Figure 5:** (a). Schottky Barrier heights of Au-Se and W-Se contact under low power illumination; (b). Photoconductivity of Au-Se-W device under under low power illumination.

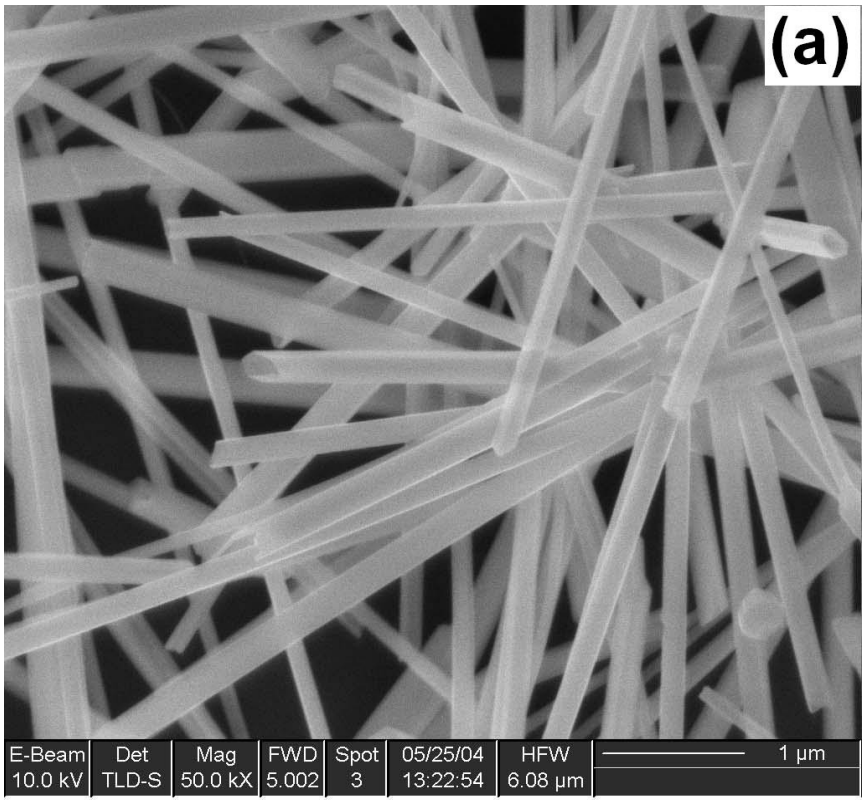
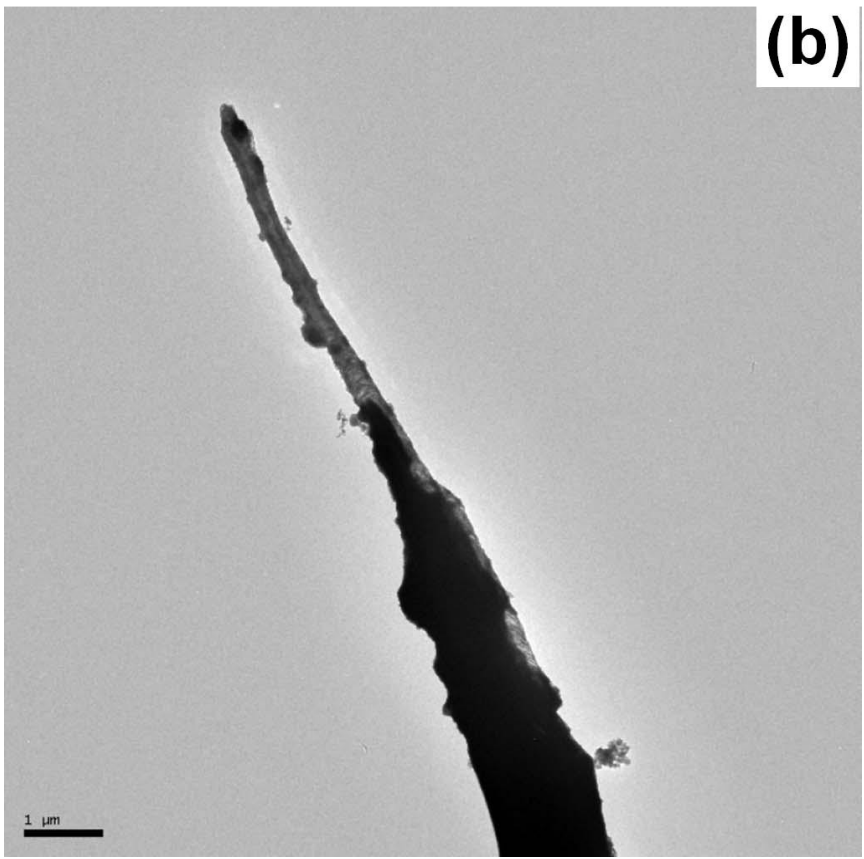

**Figure 1**

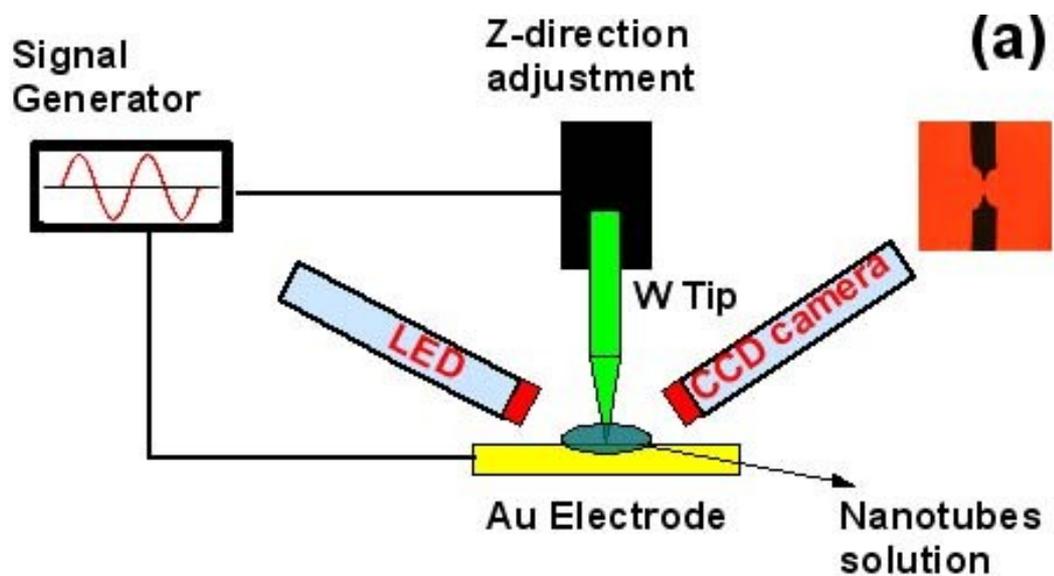

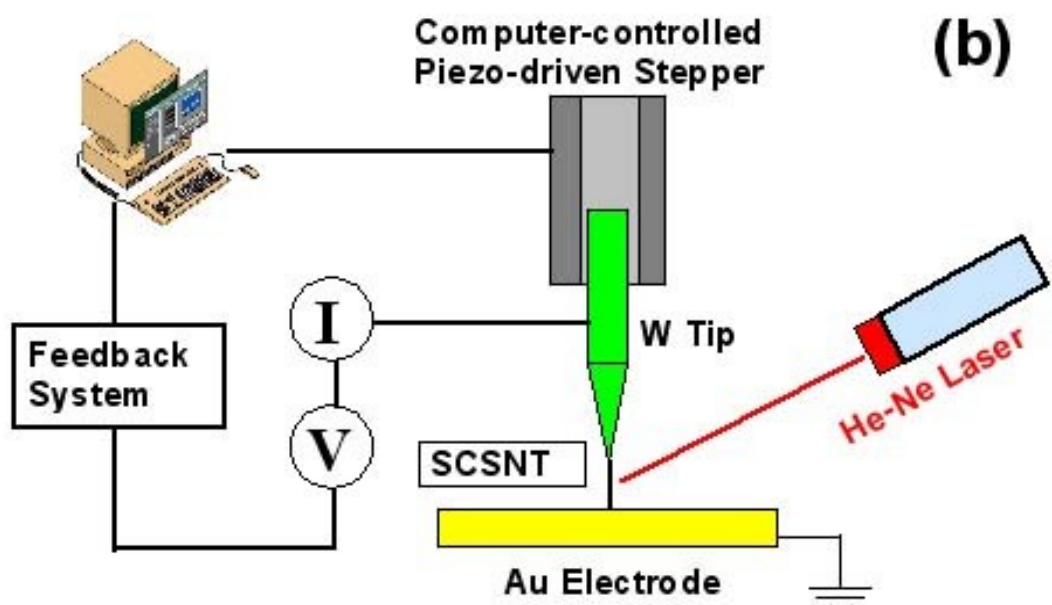

**Figure 2**

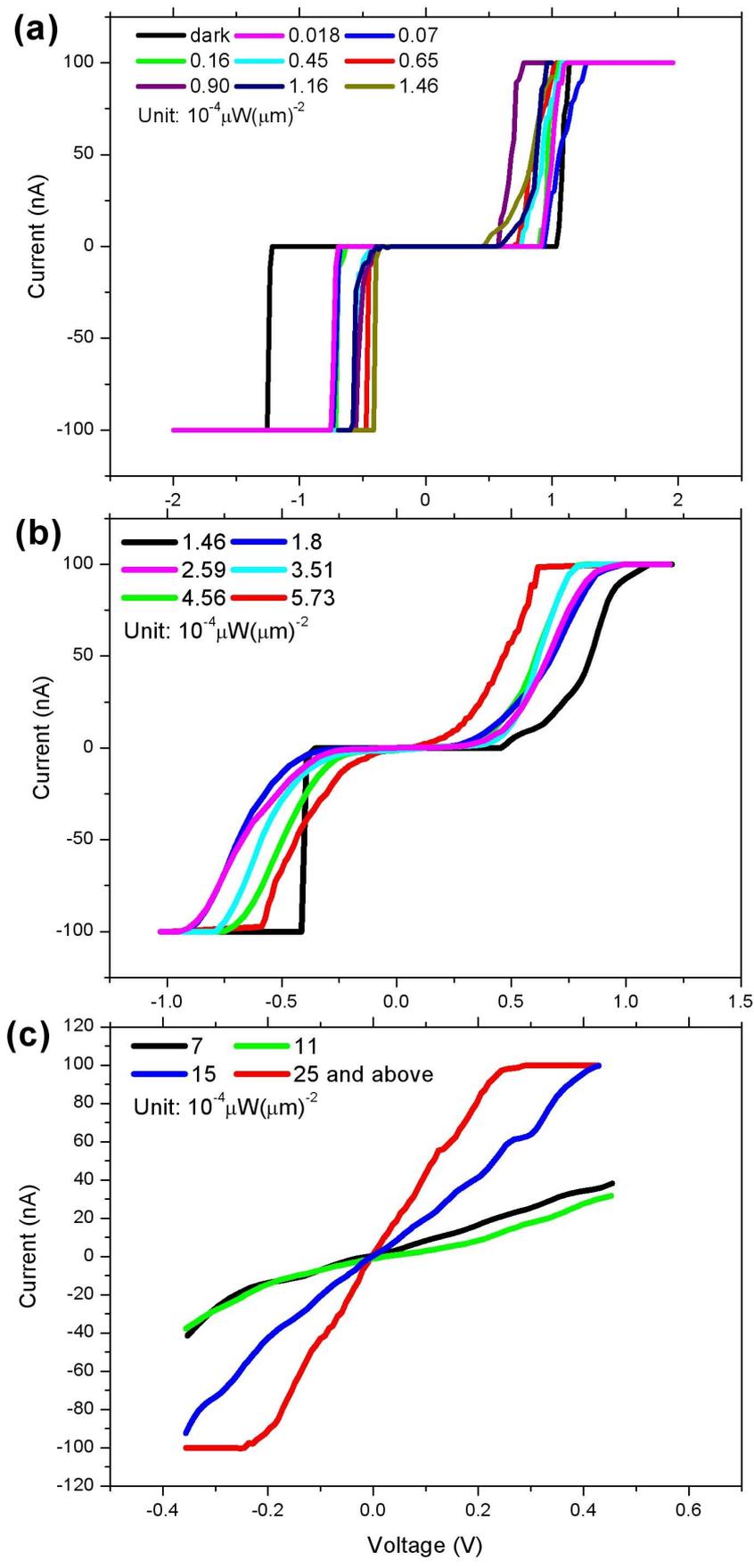

**Figure 3**

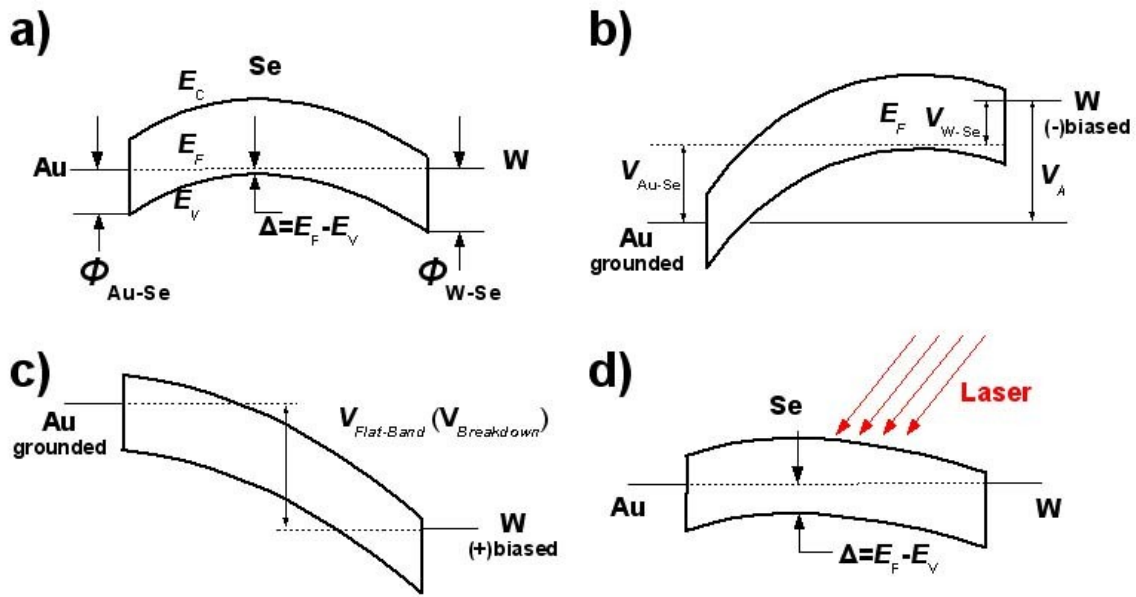

**Figure 4**

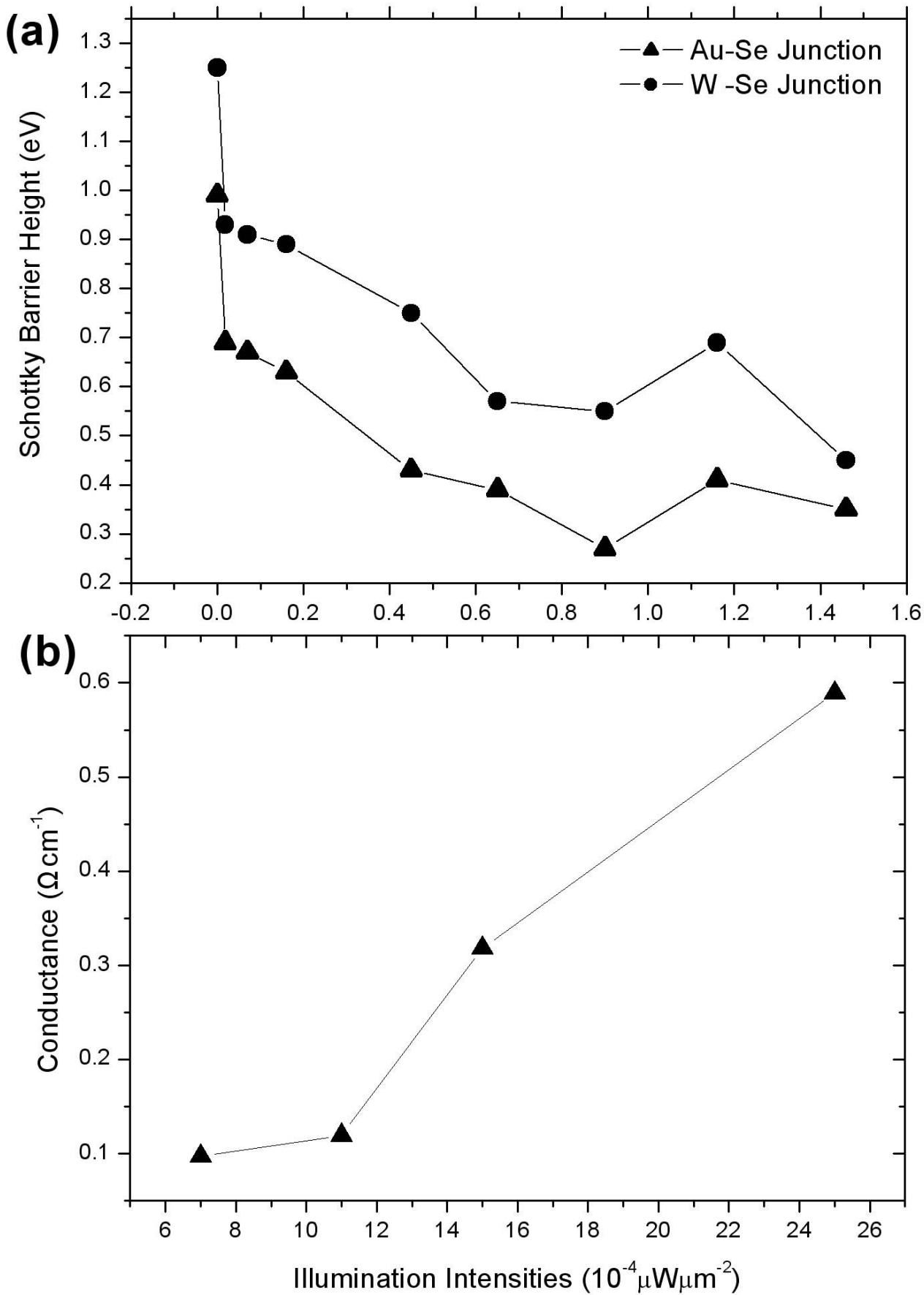

Figure 5